# Introduction to the Special Issue on Genome Organization: Experiments and Modelling

Nick Gilbert[1] and Davide Marenduzzo[2]

[1]MRC Human Genetics Unit, Institute of Genetics and Molecular Medicine, University of Edinburgh, Crewe Rd, Edinburgh, EH4 2XR, UK

[2]SUPA, School of Physics & Astronomy, University of Edinburgh, Peter Guthrie Tait Road, Edinburgh, EH9 3FD, UK

**Abbreviations**

Kb – kilobases

TF – transcription factor

3C – chromosome conformation capture

min – minute

Understanding the spatial organization of chromosomes in 3D has been a key outstanding question in biology for some time. This is a crucial area of research, because the 3D organization of chromatin at the 10-100's kilobase pair (kbp) level underlies several aspects of gene regulation, and there is evidence that it plays an important role, for example, in development (Sproul et al. 2005), aging (Pal and Tyler 2016), as well as in a number of genetic diseases (Misteli 2010).

Very recently, we have witnessed a dramatic and unprecedented rise in the number of contributions on this topic. This burst in activity is due to at least a two-fold reason. First, the development of chromosome conformation capture techniques, and especially of the high throughput variants "Hi-C", "CaptureC" and "Capture-HiC" (Osborne and Mifsud), have provided us with an impressive high-resolution catalogue of chromosomal contacts in different cell types, tissues and organisms, for healthy, senescent and diseased cells. Originally this data was at low resolution but the lower cost of sequencing has enabled large datasets to be generated providing a structural framework for genome organization in different cell types (Rao et al. 2014; Mifsud et al. 2015). Second, the refinement of polymer and statistical physics models for DNA and chromatin have made it possible to simulate the stochastic organization of genomic loci, or even entire chromosomes. These simulations are informed by existing data, but they can in turn stimulate further experiments via their predictions. "Inverse" models (some of these are reviewed by (Zhan et al.) and by (Bianco et al.)) start from the Hi-C data and work backwards, using sophisticated fitting procedures to infer a plausible polymer model; the model can then be used to make predictions on future experiments, where, for instance, the genomic region of interest is edited. "Direct" models (some of these are reviewed by Bianco et al., and by Haddad et al. 2017) start instead from simple biological and biophysical assumptions to deliver computer simulations whose output can be compared directly to Hi-C contact maps (as well as other experiments); their use is instrumental, combined with experimental evidence, to provide mechanistic models for genome organization.

There are currently two main popular models for the organization of chromosomes within 3D nuclear space: both have been prompted by the combination of insights from experimental evidence and from computer simulations. The first one assumes the organization is driven by transcription factors binding to active and inactive regions of the genome (Brackley et al. 2016b; Bianco et al. 2017; Haddad et al. 2017), whereas the second assumes that the main organizers are cohesin and condensin (Finn et al. 2017; Kalitsis et al. 2017). Both the models can explain some key aspects of genome organization, but neither can rationalize all observations.

The transcription factor (TF) model is at the basis of the "strings-and-binders" (Bianco et al. 2017) and the "block copolymer" (Haddad et al. 2017) models. The underlying idea is that chromatin conformations arise as a result of the action of bivalent (or multivalent) factors (the "binders") which can bind the chromatin fiber (the "string") at multiple points, thereby forming chromosome bridges which stabilize genomic contacts. Chromatin regions bearing active and inactive marks recruit different kinds of proteins: for instance, inactive heterochromatic regions rich in H3K9me3 bind HP1, whereas active regions (rich in H3K4me3, H3K4me1 or H3K27ac) bind holoenzymes, polymerases and transcription factors. There is evidence for the ability of the "binders" to form bridges: HP1 is known to be multivalent, and this is also the case for other repressing factors (such as PRC1 and other polycomb-group proteins), while complexes of transcription factors and polymerases will also normally have multiple DNA-binding sites. Therefore, the TF model is based on broadly valid assumptions, and it can naturally explain the segregation between euchromatin and heterochromatin (Nishibuchi and Dejardin 2017), as well as the organization of the genome into A (active) and B (inactive) compartments. Because active and inactive factors have a generic tendency to cluster (through the "bridging induced attraction", (Brackley et al. 2013), the TF model further provides a natural framework to capture the biogenesis of nuclear bodies. The TF model is appealing because it relies on minimal input (the binding sites of active and inactive factors) and in principle no fitting, yet it delivers contact maps which are in good quantitative agreement with experiments (Brackley et al. 2016b).

The TF model also relates well to studies indicating that the local transcriptional environment impacts on structure and orchestrates chromosome organization. Hi-C like techniques provide a structural basis for the genome, but appear to be relatively intransigent to transcription. Instead, a superimposition of topological data (Naughton et al. 2013) to Hi-C maps provides an additional level of domain-like organization to reveal the formation of over-wound and under-wound 100 kb chromatin domains. These correspond well to high-resolution Hi-C maps (Rao et al. 2014) and indicate the genome is organized into structural domains subdivided into functional domains regulated by transcription and topoisomerase activity. Our own simulations of these phenomena highlight how supercoiling and the implicit binding of transcriptional regulators can both control transcriptional activity and facilitate domain remodeling (Brackley et al. 2016a).

The main drawback of the TF model is that it cannot easily account for the striking observation, made through Hi-C, that chromosome loops between convergent CTCF binding sites are abundant, while those between divergent ones are virtually absent (Rao et al. 2014). This is because, at least in its simplest version, this model works in thermodynamic equilibrium, and under this condition convergent and divergent loops share the same chromatin structure: as a result, the experimentally observed bias is inexplicable within this framework. There are also some outstanding open questions which the TF model prompts. For instance, the model works well given the 1D epigenetic patterning of histone modifications, as this is a good proxy for the binding landscape of bridging factors. But how is this 1D patterning set up in the first place, and how can it be changed reproducibly during development? The "living chromatin" model outlined in (Haddad et al.) considers a chromatin fiber where the epigenetic marks can be dynamically written and erased, and may provide the right avenue to quantitatively address this question in the future (Michieletto et al. 2016).

As anticipated, the second, currently popular, mechanistic model for chromatin organization instead assumes that cohesin and condensin are the main players. Condensin has long been known to be crucial for mitotic chromatin organization (see the review by Kalitsis et al.); the idea that cohesin is fundamental to organization during interphase is the basis of the "loop extrusion" (LE) model (Fudenberg et al. 2016), reviewed by Finn et al. Like condensin, cohesin is a DNA-binding protein which topologically embraces DNA or chromatin upon binding, and stabilizes chromosome loops (Finn et al. 2017). Details of the chromatin-bound structure are still debated: a single cohesin monomer may embrace two chromatin fibers, or only one, in which case looping requires dimerization to form a molecular "hand-cuff" (Finn et al. 2017). Independently of these microscopic details, the LE model assumes that cohesin has a motor activity, which is either intrinsic or extrinsic to it, and which allows it to extrude progressively larger loops. These loops grow until they are halted by a boundary element, most likely a bound CTCF, which is known to interact with cohesin in a directional way (Fudenberg et al. 2016). The main strength of the LE model is, therefore, that it can naturally explain why almost all CTCF-mediated loops are convergent, and virtually none are divergent: this is because as the loop grows it can "sense" the orientation of CTCF, because cohesin will only stall when it faces two convergent and occupied CTCF binding sites.

The main problem of the LE model is that it assumes that cohesin actively creates loops of hundreds of kilo-base-pairs (the typical size of CTCF-mediated loops), without dissociating. Existing experimental evidence suggests that the residence time of cohesin on DNA is about 20 minutes, so, in order to extrude a loop of 100 kbp, cohesin would have to move at about 5 kbp/min, an impressive speed as this is about five times as fast as an RNA polymerase. Yet, there is not currently evidence of any motor activity of cohesin associated with unidirectional motion, as postulated in the loop extrusion model. The primary outstanding question prompted by this model (Finn et al. 2017) is therefore, what is the dynamics of cohesin-mediated chromatin loops? How does cohesin translocate on chromatin, and how fast can it do so? While the sliding of a cohesin ring embracing a single DNA molecule has recently been studied (Stigler et al. 2016), it is now necessary to characterize mobility on chromatinized DNA, and ultimately probe the dynamics of cohesin-mediated loops.

One view is that the TF and LE models are competing ones, and that there is a single main organizer, either transcription factors or cohesin/condesin, and new research will tell which one it is. Another possible view is, however, that the TF and LE models are instead complementary: after all, the first one explains A/B compartments, while the second one explains the formation of convergent CTCF loops.

Further research will then be needed to understand how the two kinds of organizers couple when they are active at the same time: for instance, are the two organizations independent of each other, or is there cross-talk between them?

Our discussion has so far been centered on large scale (10-100 kbp) features of genome organization as explored by Hi-C, where computational models have proved very beneficial in facilitating interpretation of existing experiments and also stimulated the design of new ones. Computer-based modeling is, however, possibly even more important at smaller scale of chromatin organization, down to the single nucleosome level. A different approach to study genome organization has recently been further developed by the Greenleaf lab (Risca et al. 2016): it utilizes ionizing radiation-induced spatially correlated cleavage of DNA and sequencing (RICC-seq), to provide information about local (50 – 500 bp) nucleosomal interactions. At present this data is relatively low resolution but after significant computational analysis, it gives evidence for a two-start helical chromatin fibre in heterochromatic regions of the genome but more disrupted fibres in regions of the genome with more open chromatin (Gilbert et al. 2004). These structures resemble those predicted by mesoscale modelling of chromatin folding at the scale of a few nucleosomes, where compaction is induced by proteins such as linker histones (Luque et al. 2014); this mesoscale modelling can then be further scaled up to simulate chromatin loops (Bascom et al. 2016).

In the near future, we expect that the combination of experimental and simulation techniques, which is the central theme of this Special Issue, will prove more and more effective at addressing outstanding open questions such as those we have outlined above. Such a combination has the potential to yield a transformative tool in the field, because the two approaches have different strengths and weaknesses, hence are highly complementary and can be used to ask questions which could not be answered by using either modelling or experiments alone.

Figure legend

(A) Contact maps from simulations (left, Brackley et al. 2016b) and Hi-C experiments (right, Rao et al. 2014), for chromosome 19, in GM12878 cells. Simulations were performed with a variant of the TF model, which was solely based on chromatin state and GC content (see Brackley et al. 2016b for details). (B) Simulation snapshots of the structure of chr19, with (left image) or without (right image) chromatin. Black and red proteins are inactive and active bridges respectively. For chromatin beads, red beads denote regions with strong enhancer and promoters, green beads denote transcribed regions, gray beads denote heterochromatin, and blue beads are regions which are not binding to any of the bridges (see Brackley et al. 2016b for more details of interactions between proteins and coloured chromatin beads).

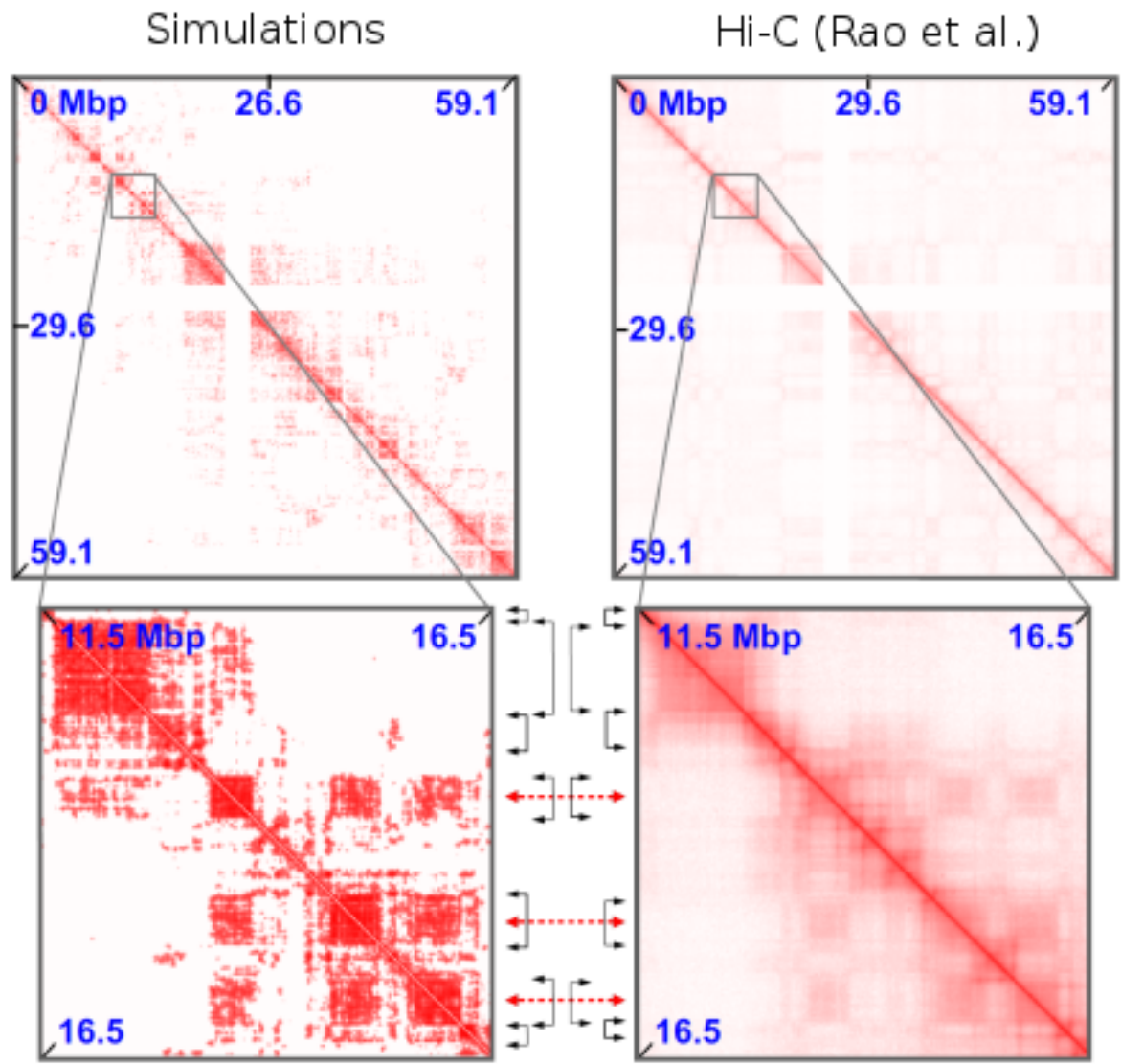
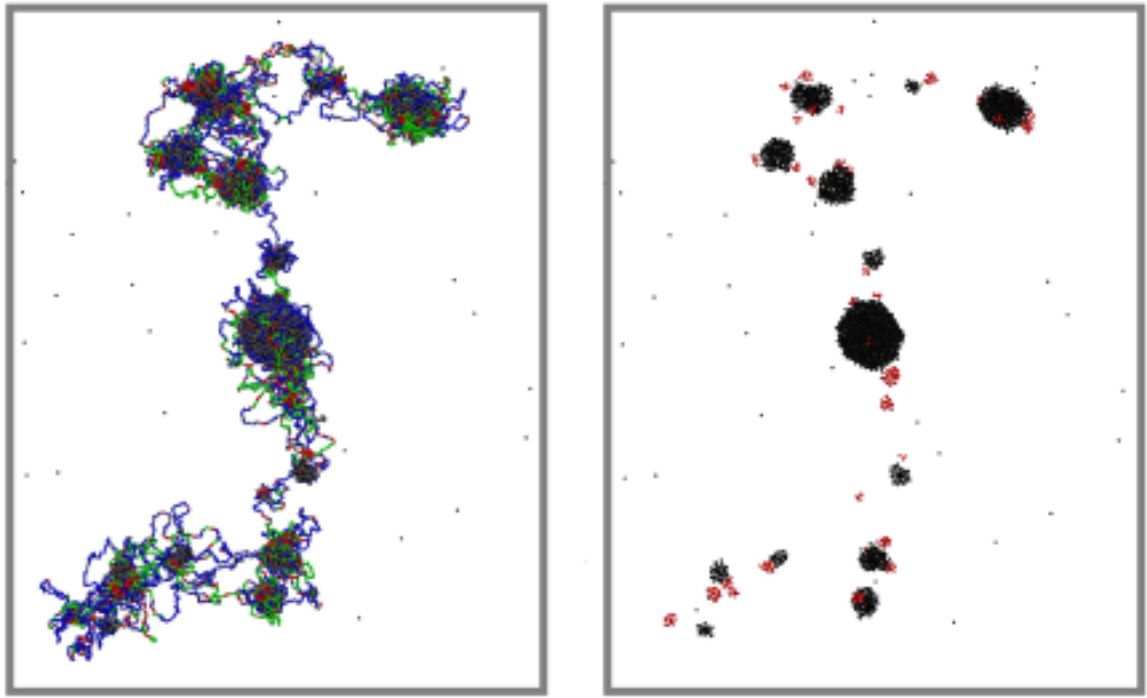